\documentclass[useAMS,usenatbib]{mn2e}

\usepackage{times}

\input{epsf.tex}

\title[Isolating integrals of an orbit]{Finding how many isolating integrals of
motion an orbit obeys}
\author[D. D. Carpintero]{D. D. Carpintero$^{1,2}$\thanks{E-mail:
ddc@fcaglp.unlp.edu.ar}\\
$^{1}$Fac. de Cs. Astron\'omicas y Geof\'{\i}sicas, Univ. Nac. de La Plata,
Paseo del Bosque S/N, 1900 La Plata, Argentina\\ 
$^{2}$Instituto de Astrof\'{\i}sica de La Plata, CCT La Plata - CONICET/UNLP,
Argentina}

\begin{document}

\date{}

\pagerange{\pageref{firstpage}--\pageref{lastpage}} \pubyear{}

\maketitle

\label{firstpage}

\begin{abstract}

The correlation dimension, that is, the dimension obtained by computing the
correlation function of pairs of points of a trajectory in phase space, is a
numerical technique introduced in the field of nonlinear dynamics in order to
compute the dimension of the manifold in which an orbit moves, without the need
of knowing the actual equations of motion that give rise to the trajectory. This
technique has been proposed in the past as a method to measure the dimension of
stellar orbits in astronomical potentials, i.e., the number of isolating
integrals of motion the orbits obey. Although the algorithm can in principle
yield that number, some care has to be taken in order to obtain good results. We
studied the relevant parameters of the technique, found their optimal values,
and tested the validity of the method on a number of potentials previously
studied in the literature, using the SALI, Lyapunov exponents and spectral
dynamics as gauges.

\end{abstract}


\section{Introduction}

The characterization of the orbits supported by an astrophysical potential is of
interest in order to study self-consistent models of stellar systems. 
\citet{s79, s82}, for example, pioneered the construction of steady state
distribution functions from a well chosen set of regular orbits belonging to a
given potential; other studies followed on the same line \citep[e.g.][]{crd00}.
Thus, it was established that regular orbits act as a dynamical skeleton of a
stellar system. Moreover, among regular orbits, those that are resonant and
stable are of utmost importance, since they generate entire families of orbits
around them: they constitute the backbone of the system. On the other hand,
chaotic orbits, the existence of which in realistic models of galaxies is
nowadays beyond doubt \citep{vm98,vks02,mcw05}, are important to the dynamical
evolution of stellar systems. In particular, the slow difusion of many chaotic
orbits through their allowed phase space may affect even the global
characteristics of the system. As \citet{m03} and \citet{mm04} showed, the
spatial distribution of fully chaotic orbits (i.e., those which obey only one
isolating integral of the motion) and partially chaotic orbits (those which obey
two integrals) is quite  different. This points towards a different dynamical
role for each type of chaoticity, although what kind of role is still unknown.

Thus, finding regular, resonant, non resonant, partially chaotic and fully
chaotic orbits is fundamental in the study of the dynamics of a stellar system.
They all are defined by the number of isolating integrals of motion that they
obey. For an $N$-dimensional potential, regular orbits have $N$ or more
isolating integrals, whereas chaotic orbits obey less than $N$ isolating
integrals\footnote{A regular orbit, by definition, has $N$ or more isolating
integrals of motion. However, a chaotic orbit is defined through its sensitivity
to the initial conditions in phase space: if the initial conditions of the orbit
are infinitesimally displaced, the distance between the original orbit and the
new orbit grows exponentially with time. These definitions do not complement
each other. Whereas it can be proved that a regular orbit is not chaotic and a
chaotic orbit is not regular \citep[e.g.][\S 8.3]{j91}, it has not been proved
that every irregular (i.e. not regular) orbit is chaotic, or, in other words,
that every orbit obeying less than $N$ isolating integrals has sensitivity to
the initial conditions. Nevertheless, to avoid confusion, we will follow here
the widespread convention of considering irregular orbits and chaotic orbits as
the same set.}. Among regular orbits, those that are not resonant obey exactly
$N$ isolating integrals, whereas those resonant have one more isolating integral
for each additional resonance. On the other hand, among chaotic orbits, fully
and partially chaotic orbits are also distinguished by the number of isolating
integrals they have, as said above. Thus, a method allowing to determine the
number of isolating integrals an orbit obey is a fundamental tool in studying
the dynamics of a stellar system. 

One of the main consequences of obeying isolating integrals of motion is the
dimension of the manifold on which the orbit moves. Since an isolating integral
is, by definition, a non-degenerate\footnote{A non-degenetate function takes, at
most, a finite or infinite countable number of values for each set of values of
its independent variables. This is the condition distinguishing isolating and
non isolating integrals of motion.}, time independent funcion of the phase space
coordinates the value of which is constant along the orbit, it reduces in one
the dimension of the manifold in which the orbit moves \citep[][ \S 3]{bt08}.
Thus, if $N_{\rm ps}$ is the dimension of the phase space, an orbit obeying
$N_{\rm i}$ isolating integrals moves in a manifold of dimension $N_{\rm
ps}-N_{\rm i}$. One can then ascertain the number of integrals that an orbit
obeys by computing the dimension of the space in which it moves. To be brief, we
will refer hereafter to 'isolating integrals' as simply 'integrals', and to 'the
dimension of the manifold on which an orbit moves' as 'the dimension of an
orbit'.

The traditional methods to compute the dimension of an orbit or the number of
its integrals are: a) Surfaces of section \citep[e.g.][]{hh64,c83}. As is well
known, this method needs a qualitative judgment on a plot, and cannot
distinguish between orbits moving in 3 o more dimensions. b) Lyapunov exponents
\citep[e.g.][]{bggs80}. This method is the standard tool to separate regular
from chaotic orbits; it is quite reliable, and can in principle recognize the
dimension of chaotic orbits. On the negative side, since one needs both to
integrate a set of variational equations and to integrate along large intervals
in order to approximate $t\to\infty$, it is quite expensive in terms of
computing time, and, furthermore, it cannot recognize the dimension of regular
orbits. c) Spectral dynamics \citep[e.g.][]{bs82, ca98}. Taking into account
that a regular orbit on an $N$-dimensional potential moves on a manifold
diffeomorphic to a $N$-dimensional torus \citep{a89}, the natural frequencies of
revolution around the $N$ independent circles of the torus must be reflected in
the Fourier spectra of the orbit. This allows to find whether there are
resonances, and thus the dimension of the orbit can be established.
Unfortunately, the method is not suitable for analizing rotating potentials,
requires a fine tuning of a set of numerical parameters for each different
potential, and cannot find the dimension of chaotic orbits. The frequency map
method \citep{l90} is based on the same idea. d) The Smaller Alignment Index, or
SALI \citep[e.g.][]{sabv04}, and its generalization, the Generalized Alignment
Index, or GALI \citep[e.g.][]{sba07}. Both methods are based in a geometrical
property of phase space vectors joining close initial conditions, namely that
they align themselves or not depending on the geometrical properties of the
local dynamics of the system, allowing a fast and accurate determination of
whether an orbit is regular or chaotic. The GALI, in particular, although cannot
retrieve the dimension of a chaotic orbir, do allow to discern the dimension of
the torus of regular orbits, making it an excellent gauge to our results.

\citet{gp83} have proposed a method to find the dimension of (the attractor of)
an orbit in an arbitrary dynamical system, the equations of motion of which are
unknown, based on the computation of the correlation integral of the time series
of an arbitrary observable. In this context, several improvements have been made
to the method \citep[e.g.][]{dk90, dgosy93}, and many caveats and spurious
estimations have been found \citep[e.g.][]{ks97}. Also, some implementations
have been made in the astronomical field \citep[e.g.][]{hlvc98}. The application
of the method to find the dimension of an orbit integrated in an astronomical
potential, i.e., knowing the equations of motion and being able to sample the
actual trajectory of the system at arbitrary points, was first proposed by 
\citet{cs84}, and more recently by \citet{b01}.  The rationale is as follows.
Suppose that a given orbit obeys $N_{\rm ps}-1$ isolating integrals, i.e., it is
closed and therefore unidimensional. Given a small hypersphere (in phase space)
of radius $r$ around a point $P$ of this orbit, the number of other points of
the orbit included into the sphere will grow linearly as $r$ grows. If the orbit
obeyed $N_{\rm ps}-2$ integrals, moving in a bidimensional surface, the number
of points would grow as $r^2$ as $r$ grows. From this point on, it is clear that
if the orbit is moving on a $D$-dimensional space, the number of points would
grow as $r^D$ as $r$ grows. This simple mechanism, in principle, allows to
easily compute the dimension of the orbit and, therefore, the number of
integrals of motion it obeys.

\section{The method and its drawbacks}

\citet{cs84} and \citet{b01} have developed a simple algorithm in order to
compute the dimension $D$ of an orbit following the foregoing idea. Given a set
of phase space points of the orbit, first compute the distances between pairs of
them. Second, compute the cumulative distribution function of those distances.
This is nothing but the correlation integral $C(r)$ of the points of the orbit,
i.e., an histogram of the number of pairs of points separated by a given
distance or less. This function must be precisely equal to $r^D$, i.e., the
number of pairs closer than a given distance must grow as $r^D$. Therefore, from
a log--log plot of $C(r)$, a straight line can be fitted, its slope giving the
desired exponent $D$ which is the dimension sought. A few caveats are in order:
a) The exponent $D$ is theoretically obtained only in the limit $r\to0$; thus,
the slope must be computed avoiding large values of $r$. b) Unfortunately, since
the volume of phase space is proportional to $r^{N_{\rm ps}}$, there will be
always few points in the region $r\simeq0$, so this region of $C(r)$ will be
noisy in general, and therefore must be also avoided in computing $D$. c) In
order to avoid autocorrelations, the points of the orbit should not be too close
in time; otherwise, $C(r)$ will tend to yield $D=1$ at short distances (see \S
\ref{timestep}). Therefore, in order to build up the histogram, the distances
should be computed only between a random subsample of the integrated points. Of
course, the number of points $N_{\rm p}$ of the orbit must be sufficient so as
to get a reasonable good computation of $C(r)$. 

\begin{figure}
\epsfxsize=240pt\epsfbox{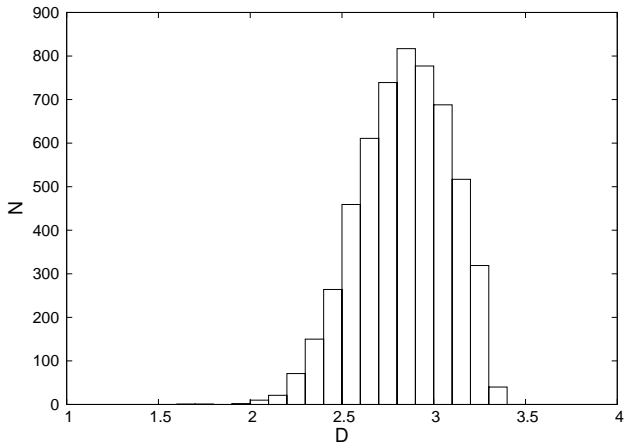}
\caption{Number of orbits $N$ with dimension $D$, for 5487 orbits integrated in
a St\"ackel potential, using $10^4$ points out of the $10^5$ points of the orbit
to compute the correlation integral.}
\label{DStackel}
\end{figure}

\begin{figure}
\epsfxsize=240pt\epsfbox{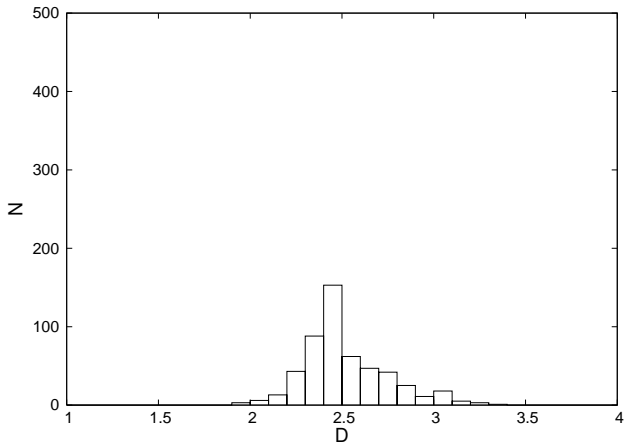}
\caption{Same as Fig. \ref{DStackel}, but only for those orbits that had $D\le
2.5$ in the original classification, and here using $10^5$ points out of the
$10^6$ points of the orbit  to compute the correlation integral.}
\label{DStackel2}
\end{figure}

As described, the method is very simple indeed. \citet{cs84} and \citet{b01}
give several examples of its application to orbits supported by different
astronomical potentials, and the results seem encouraging. However, in order to
compete against other methods, the algorithm must be capable of classify orbits
blindly; otherwise, it would be not better than inspecting surfaces of section.
To see whether it is the case, we choose a St\"ackel potential, for it has all
its orbits regular (i.e., they all must have $D=3$, with the possible exception
of a few orbits with $D=2$), and therefore we know the correct outcome
beforehand, allowing us to assess immediately how good is the method, at least
in this particular case. Thus, we integrated 5487 orbits in the potential
generated by the perfect ellipsoid density distribution 
\begin{equation}
\rho({\bmath r})={\rho_0 \over (1+m^2)^2},
\label{densidad}
\end{equation}
where
\begin{equation}
m^2={x^2\over a^2} + {y^2\over b^2} + {z^2\over c^2},
\end{equation}
and $\rho_0$, $a\le b\le c < 0$ are constants \citep{d85}. We used $a=-1$, $b=
-0.390625$, $c=-0.25$, and $\upi^2\rho_0 abc=1$. The orbits were integrated
along 500 periods with a time step of 1/200 of the respective orbital period, so
as to get 100,000 points for each orbit. The dimensions $D$ were computed using
the distances of a random sample of 10,000 of those points. The result is shown
in Fig. \ref{DStackel}, where it can be seen a wide dispersion of values instead
of the unique expected value $D=3$. In order to verify whether the number of
points $N_{\rm p}$ was inadequate \citep[][\S 4]{b01}, we took the 520 orbits
that lie to the left of $D=2.5$, and integrated them along 5000 periods,
effectively increasing both the number of orbital points and the number of
sample points by ten. The result is showed in Fig. \ref{DStackel2}, where it can
be seen again that the orbits do not pile up around $D=3$. This result, besides
showing that a na\"{\i}ve implementation of the method does not yield good
results, highlights another point: a non-integer value of $D$ does not
necessarily denote a fractal dimension \citep{cs84}, but may indicate a poor
computation of the true dimension.

\section{Numerical implementation}

It is clear that a careful numerical implementation of the method is essential
to get acceptable results. We have identified several parameters that affect the
outcome. To illustrate the analysis, we will use, in the following, two orbits
integrated in the St\"ackel potential generated by the perfect ellipsoid density
distribution (Eq. [\ref{densidad}]). Table \ref{tabla1} shows the respective
initial conditions. The first orbit is a box orbit; the second one is a
$z$--tube orbit; both orbits move in a manifold of dimension $D=3$. Also, in all
cases the function $C(r)$ has been normalized to its maximum value, that is,
$C(r)=1$ at large distances.

\begin{table}
\centering
\caption{Initial conditions of the example orbits.}
\label{tabla1}
\begin{tabular}{@{}lrr@{}}
\hline
      & Box                    & $z$--tube             \\
\hline
$x$   & $1.163562178611755$    & $0.5087397098541260$  \\
$y$   & $-0.8299766778945923$  & $-1.472079634666443$  \\
$z$   & $0.06499148160219193$  & $0.8406194448471069$  \\
$v_x$ & $0.2304639220237732$   & $0.4684486091136932$  \\
$v_y$ & $0.1688854843378067$   & $0.01424702629446983$ \\
$v_z$ & $-0.04289411380887032$ & $-0.1927715390920639$ \\
\hline
\end{tabular}
\end{table}

\subsection{The window}

\begin{figure}
\epsfxsize=240pt\epsfbox{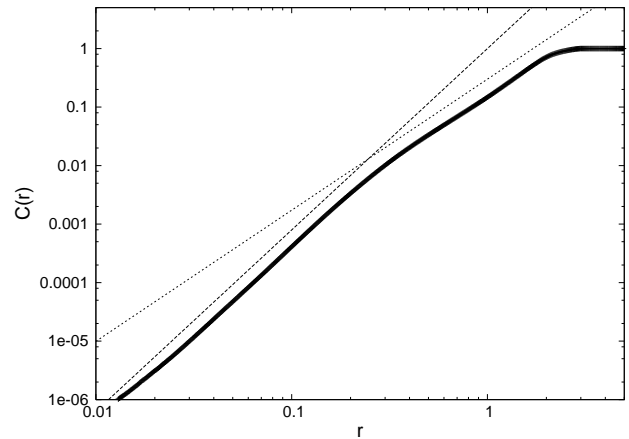}
\caption{Correlation integral $C(r)$ of the box orbit. The straight lines have
slopes 3.10 and 2.24, respectively.}
\label{dospend}
\end{figure}

In order to automate the procedure, the portion of $C(r)$ to be used to fit the
straight line --the {\sl window}-- must not be chosen by eye.  We solve the
problem by using a mobile window, that is, by computing the slope of $C(r)$
several times, each one using only the points defined by a window of fixed
length placed in a different location of the function $C(r)$. In order not to
miss any good fitting, the window is placed starting at every computed point of
$C(r)$. Also, due to the uncertainty of the slopes computed with too few points,
the length of the window has to have a minimum; a window spanning at least a
decade in $r$ should suffice \citep{gp83}. Once the slopes have been computed
for each position of the window (using least squares fitting), we choose as the
desired slope that with the minimum error. Here we define 'error' as the value
of $\chi^2$ yielded by the least squares fitting \citep{ptvf94}. But, as Fig.
\ref{dospend} shows, there is still a problem to solve: the correlation integral
typically admits two possible slopes, $D=3.10$ and $D=2.24$ for the case of the
box orbit of the Figure. 
\begin{figure}
\epsfxsize=240pt\epsfbox{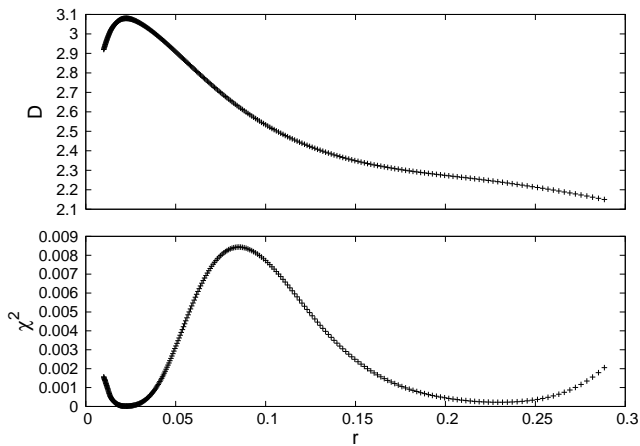}
\caption{Upper panel: dimension $D$ obtained from least squares fittings to the
function $C(r)$ illustrated in Fig. \ref{dospend}. The abscisae of each point
corresponds to the starting point of a decade-long window used to obtain the
corresponding value of $D$. Lower panel: the error $\chi^2$ for each value of
$D$. Two minima can be clearly seen; the left one is always chosen to compute
$D$, whenever the neighboring points reach enough height.}
\label{Dsji2}
\end{figure}
Fig. \ref{Dsji2} shows the error of the fitting as the window was displaced
along the function $C(r)$, and the value of the corresponding slope. As can be
seen, the error has two minima, corresponding to the two slopes seen by eye.
Which minimum should be chosen? To answer this, it has to be taken into account
that the correlation integral gives the dimension of the orbit only at small
distances, and, moreover, that at larger values of $r$, the cumulative function
$C(r)$ saturates at two points: a) when all the distances have been incorporated
into the histogram ($C(r)=1$), and b) well before that, when the distances are
comparable to the typical diameter of the volume of phase space occupied by the
orbit, because there is a lack of points at distances so large. Therefore, it
seems safe to claim that the slope with the minimum corresponding to small
distances is the correct one. We accomplish this by simply picking up the value
of the slope corresponding to the first minimum found when the window is
displaced from small values of $r$ towards larger values. In the example, the
correct slope $D=3.1$ is thus chosen.

\begin{figure}
\epsfxsize=240pt\epsfbox{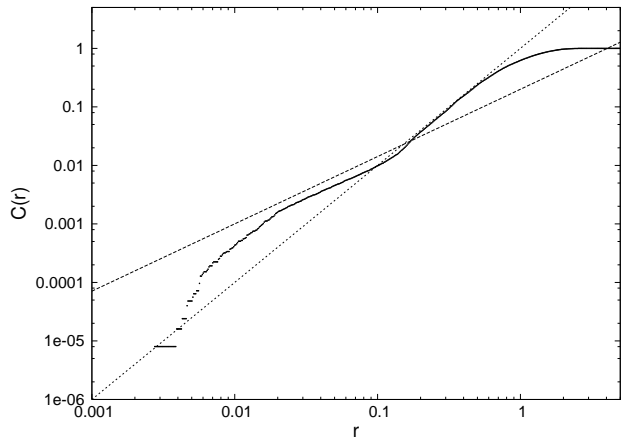}
\caption{Correlation integral $C(r)$ of an
orbit integrated in a Plummer potential $\Phi=-C/\sqrt{r^2+a^2}$ with $C=1$ and
$a=0.2945243$. The orbit is a rosette ($D=2$), but the generating ellipse
rotates at a very low angular velocity, and the orbit is seen with $D=1$ at
short distances. The straight lines have slopes 1.15 and 2, respectively.}
\label{invertidas}
\end{figure}

\begin{figure}
\epsfxsize=240pt\epsfbox{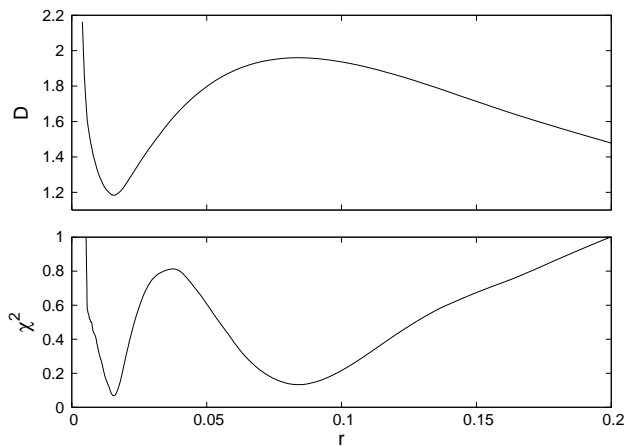}
\caption{As Fig. \ref{Dsji2}, but for the orbit of Fig. \ref{invertidas}. The
first minimum of $\chi^2$ is not shallow.}
\label{Dsinvertidas}
\end{figure}

Nevertheless, there are cases in which the first minimum does not yield the real
dimension of the orbit, These include shallow minima indicating a region of low
curvature in $C(r)$, giving smaller errors than the surrounding regions, but not
being representative of the dimension of the orbit. To avoid choosing these
minima, a minimum is taken only if the surrounding values of $\chi^2$ reach at
least three times the own value of the minimum. Still, in other cases, the first
minimum does not correspond to the true value of $D$ because it is a
circumstancial good fitting. Fig. \ref{invertidas} shows one of these cases. In
fact, in this example, a longer integration yields $D=2$, once the orbit begins
to cover densely its 2D manifold. Although the abovementioned shallow minimum
argument is applicable in many of these cases, there are examples in which it is
not. Fig. \ref{Dsinvertidas} shows that the first minimum in $\chi^2$
corresponding to the orbit of Fig. \ref{invertidas} is very deep, so the
dimension to which it corresponds ($D=1.15$) will be taken as correct. Since the
regions of these first minima often span less than one decade in $r$, one could
shorten the length of the window in order to rise the errors and turn the minima
into shallow ones; however, as said before, the shortest acceptable window must
span at least one decade in order to achieve good results in the general case.
Therefore, orbits that behave like the one described will be missclassified
unless they are integrated during longer periods.

As an additional safeguard, the values of $C(r)$ at very small distances are not
included in any window, thus avoiding the noisy region, which may introduce
spurious 'good' values of $D$ by chance. Since the information about $D$ is
contained at small values of $r$, this cut should be chosen wisely; it should
not embrace any valuable region of $C(r)$. We have found that cutting the region
$C'(r) < 100$ from the analysis, where $C'(r)$ is the unnormalized correlation
integral, effectively avoids the noisy region without affecting the results.

\subsection{The number of points to compute distances}

\begin{figure}
\epsfxsize=240pt\epsfbox{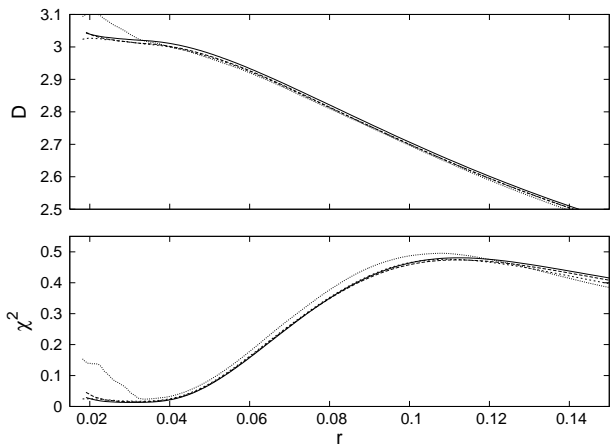}
\caption{Top: Slopes of the different portions one decade long of the
correlation integral of the $z$--tube orbit when computed with $f=0.02$ (dotted
line), $f=0.05$ (long dashed line), $f=0.1$ (solid line) and $f=0.2$ (short
dashed line, almost superimposed over the solid line). Bottom: errors associated
with them.}
\label{Dsji2.nm} 
\end{figure}

The main parameter is, undoubtely, the number of points $N_{\rm p}$ taken from
the orbit in order to compute distances among them and thus build up the
correlation integral $C(r)$. Let $N_{\rm orb}$ the number of integrated points
of the orbit, and let $f$ the fraction of those points used to compute $C(r)$,
i.e., $N_{\rm p}=fN_{\rm orb}$. Fig. \ref{Dsji2.nm} shows the result of applying
several values of $f$ to the $z$--tube orbit, using $N_{\rm orb}=300,000$. As
can be seen, unless $f < 0.05$ --in which case the autocorrelation of the points
of the orbit begins to play a role--, the result is robust enough. Taking into
account that the larger is $f$ the more numerically expensive is the computation
of $C(r)$, we chose the value $f=0.1$.

\begin{figure}
\epsfxsize=240pt\epsfbox{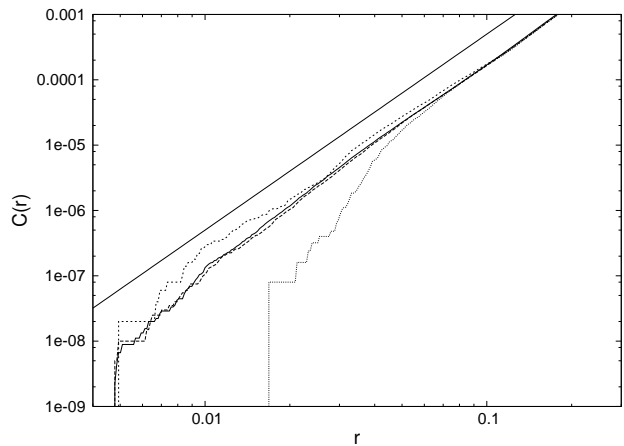}
\caption{Correlation integral of the $z$--tube orbit at small distances,
computed using different number of points $N_{\rm p}$. A straight line of slope
3 has been added for reference. Dotted line: $N_{\rm p}=5\times 10^3$,
short-dashed line: $N_{\rm p}=10^4$, long-dashed line: $N_{\rm p}=2\times 10^4$,
solid line: $N_{\rm p}=3\times 10^4$.}
\label{cder.np}
\end{figure}

\begin{figure}
\epsfxsize=240pt\epsfbox{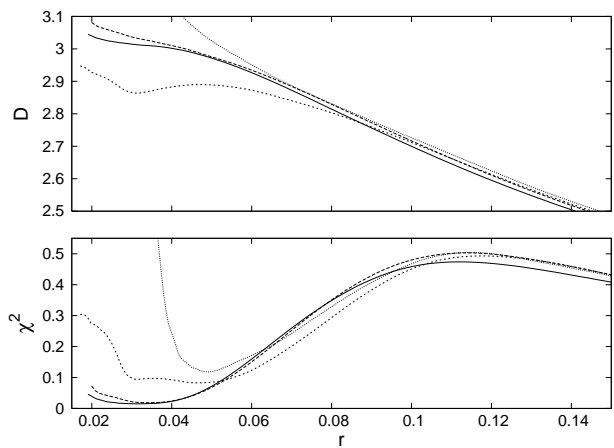}
\caption{Slopes of the different portions one decade long of the curves in Fig
\ref{cder.np}, and errors associated with them. Lines as in Fig. \ref{cder.np}.}
\label{Dsji2.np}
\end{figure}

Once $f$ is fixed, there remains the choice of $N_{\rm p}$ --which in turn
determines $N_{\rm orb}$ through $f$. Fig. \ref{cder.np} shows, for the
$z$--tube orbit, the correlation integral at small distances when $N_{\rm p}$ is
varied from $5\times 10^3$ to $3\times 10^4$. We can see that this regime of
small distances, being the most important, is quite affected by $N_{\rm p}$.
Only when $N_{\rm p} \simeq 2\times 10^4$ the slope is close to 3. Further
increases of $N_{\rm p}$ did not improve significantly the result in this
example. Fig. \ref{Dsji2.np} shows the respective slopes and errors computed for
this case.

In repeating this kind of experiment with other 3D potentials in which chaotic
orbits are allowed (the triaxial generalization of the Dehnen potential
\citep{mf96} and the numerical potential described below), we found that greater
values of $N_{\rm p}$ were required, due to the fact that the number of points
needed to cover a 4D or 5D manifold is quite large; we have found that $N_{\rm
p} = 5\times 10^4$ is enough to cope with $D=5$ orbits (i.e., $5\times 10^5$
orbital points); larger values did not, in general, improve the results. 2D
potentials, on the other hand, required fewer points; $N_{\rm p} = 3\times 10^4$
suffice even for chaotic orbits. Therefore, we adopted $N_{\rm p} = 5\times 10^4$
for 3D potentials and $N_{\rm p} = 3\times 10^4$ for 2D potentials.  

\subsection{The time step}
\label{timestep}

The integration of the orbit is an essential first step in computing its
dimension. Besides the abovementioned total number of points of the orbit, the
time step $\Delta t$ with which the coordinates are advanced is another
important parameter in order to compute a correct value of $D$. First, if
$\Delta t$ were too small, it would make the points to be too close to one
another; this in turn would make them to be biased towards dimension 1, because
there are too many points aligned along a curve; furthermore, this alignment
will recur each time the orbit enters the hypersphere. This can be avoided by
simply discarding neighbouring points, but this renders the small time step
useless. Better off is to take a larger time step. On the other hand, $\Delta t$
should not be too large: besides the longer times of integration needed to get
the desired number of points of the orbit, it may yield an insufficient number
of close points in order to render a meaningful correlation integral at short
distances. Thus, $\Delta t$ must be judiciously chosen. It is clear that a fixed
value is useless, because each orbit has its own orbital period and visits the
phase space at its own rate; a time step good enough to achieve a fair sample of
points of a given orbit into every hypersphere, may be bad for another orbit.
Therefore, it is better to fix the time step as, for example, a given fraction
of the orbital period $T_{\rm p}$. This was implemented by the following
procedure. First, we locate the coordinate origin at the baricenter of the
density distribution. Then, we integrate the orbit with an (initially small)
arbitrary time step and, by counting coordinate planes crossings, estimate its
period. The integration is then started again, now taking as a time step a given
fraction of this estimated period. This algorithm works fine for regular orbits,
which have a definite period. Chaotic orbits, on the other hand, lack in general
a definite dynamical period. Nevertheless, since the aim is to get orbital
points not too close and not to far to one another, the abovementioned procedure
still works even in the case of a not well-defined period, as long as the time
step is chosen based on an enough number of plane crossings so that a mean time
of return to the same octant can be computed.

\begin{figure}
\epsfxsize=240pt\epsfbox{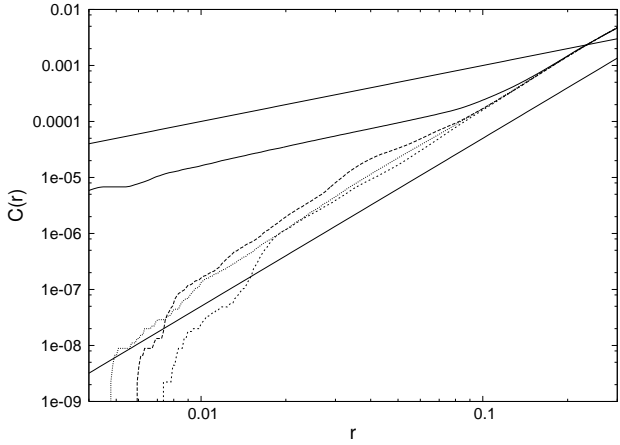}
\caption{Correlation integral computed using different time steps. Two straight
lines of slopes 1 and 3 have been added for reference. Solid line: $\Delta
t=0.0003 T_{\rm p}$; long-dashed line: $\Delta t=0.003 T_{\rm p}$; short-dashed
line: $\Delta t=0.03 T_{\rm p}$; dotted line: $\Delta t=0.3 T_{\rm p}$.} 
\label{cder.deltat}
\end{figure}

Fig \ref{cder.deltat} shows several curves  $C(r)$ corresponding to the
$z$--tube orbit, computed using different time steps; there are also two
straight lines with slopes $m=1$ and $m=3$ for reference. As expected, a very
small $\Delta t$ makes the slope to give $D=1$ at short distances. The correct
value of $D$ is obtained just when $\Delta t \simeq 0.3 T_{\rm p}$, i.e., a
considerable fraction of the orbital period. Larger time steps do not
appreciably improve the result, and, moreover, the integration becomes very
difficult to carry out. We have found in practice that $\Delta t = 0.3 T_{\rm
p}$ is a good choice.

Ideally, the time step should be further scaled proportionally to the diameter
of the volume of phase space visited by the orbit, to assure a well distributed
sea of points and therefore a well defined $C(r)$ at all distances. This was
also tried, but no appreciably improvements were observed by taking into account
this correction.

\subsection{The precision of the integration}

\begin{figure}
\epsfxsize=240pt\epsfbox{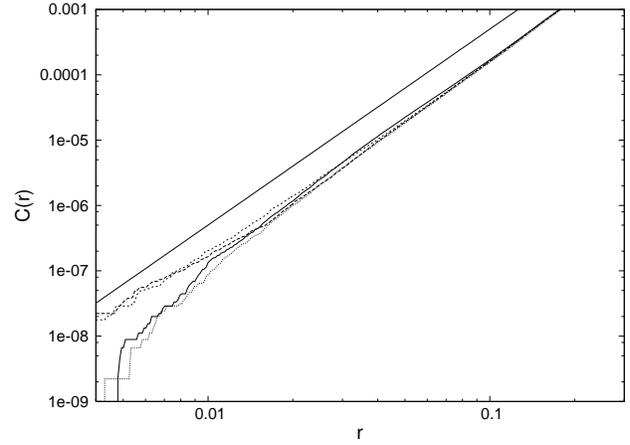}
\caption{Correlation integral of the $z$--tube orbit, computed using different
values of $\varepsilon$. A straight line of slope 3 has been added for
reference. Dotted line: $\varepsilon=8\times 10^{-3}$, short-dashed line:
$\varepsilon=8\times 10^{-4}$, long-dashed line: $\varepsilon=8\times 10^{-5}$,
solid line: $\varepsilon=8\times 10^{-6}$.}
\label{cder.prec}
\end{figure}

The precision with which the integration of the orbit is performed has also an
influence in the outcome. Let $\varepsilon\equiv |E_{\rm f} - E_0|/|E_0|$ the
relative energy error in integrating the orbit, where $E_0$ and $E_{\rm f}$ are
the energies per unit mass of the orbit at the start and at the end of the
integration, respectively. Fig. \ref{cder.prec} shows how this parameter shifts
the curve $C(r)$ of the $z$--tube orbit at small values of $r$. In general, we
have found that, whereas  $\varepsilon \approx 10^{-5}$ or better, the function
$C(r)$ maintains a correct slope. Overall, the needed precision seems not to be
too demanding; a limit value of $\varepsilon=10^{-5}$ has been adopted.

\subsection{The normalization of the phase space}

\begin{figure}
\epsfxsize=240pt\epsfbox{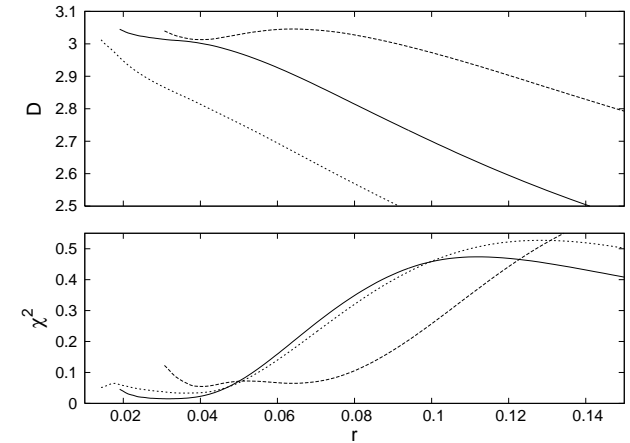}
\caption{Top: Slopes of the different portions one decade long of the
correlation integral of the $z$--tube orbit when computed with $\delta =
\sigma_x / \sigma_v$ (long dashed curve), $\delta = 1$ (solid curve), and
$\delta = \sigma_v / \sigma_x$ (short dashed curve). Bottom: errors associated
with them.}
\label{Dsji2.fase} 
\end{figure}

The normalization $\delta$ of the distance in the phase space is the factor that
should be introduced in computing a Cartesian distance $d$ between two phase
space points, due to the different nature of the positions and the velocities:
\begin{equation}
d=\sqrt{x^2+y^2+z^2+\delta^2(v_x^2+v_y^2+v_z^2)}.
\end{equation}
Since we are dealing with a correlation of distances in phase space, this factor
could be very important. We have tried several possible functional forms of
$\delta$. Fig. \ref{Dsji2.fase} shows the result of applying to the $z$--tube
orbit the normalization
\begin{equation}
\delta=\sqrt{{\sigma_x^2 \over \sigma_v^2}},
\end{equation}
where $\sigma_x$ and $\sigma_v$ are the dispersions in position and velocity of
the points of the orbit, respectively. This value is used in an attempt to
mitigate any differences between the numerical values of both positions and
velocities. It is also shown in the figure the outcome when using the reciprocal
of the abovementioned value of $\delta$, in order to try to emphasize any
underlying effect produced by the different nature of positions and velocities,
and also it is shown the result when no normalization ($\delta=1$) is used at
all. Although it is clear that the normalization indeed affects the results, the
best result, surprisingly enough, is obtained without any normalization at all.
We have found that this is, in general, the case: almost always the lack of
normalizaton yields the best slopes.

\begin{figure}
\epsfxsize=240pt\epsfbox{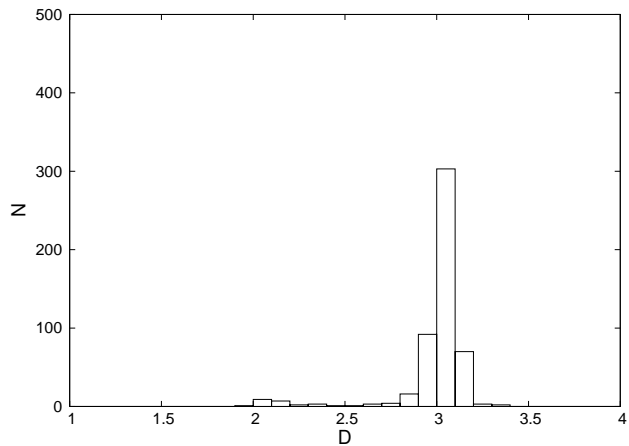}
\caption{Same as Fig. \ref{DStackel2}, but using the numerical values of the
parameters analysed in the text.}
\label{DStackel3}
\end{figure}

We reproduced the experiment shown in Fig. \ref{DStackel2}, but using the
numerical values found along this section. Fig. \ref{DStackel3} shows the
result. As can be seen, almost all orbits now pile up around $D=3$. A few
orbits, however, are still far from $D=3$. Some of them can be made $D=3$ orbits
by increasing $N_{\rm p}$; some others by increasing $\Delta t$. This shows a
fundamental limitation of the method: as in the case of the computation of
Lyapunov exponents, there are orbits that demand very long integration times in
order to be correctly classified.

Having established acceptable values of the main parameters in the case of an
integrable 3D potential, there remains the issue of verifying that those values
are indeed good enough when applied to other cases: 2D potentials, non
integrable potentials, or both. We therefore computed the dimensions of a set of
orbits integrated in the 2D, non integrable H\'enon--Heiles potential
\citep{hh64}
\begin{equation}
\Phi_{\rmn H}= {1\over 2}\left(x^2+y^2+2x^2y-{2\over 3}y^3\right)
\label{phenon}
\end{equation}
at an energy $E=0.125$, varying the foregoing parameters, using the results
obtained with the SALI indicator \citep{sabv04} as a gauge. In this last case,
we integrated the orbits until $t=1000$ units, and considered an orbit as
regular whenever SALI$>10^{-2}$, and chaotic otherwise. Table \ref{tabla3} shows
the percentages of coincidence between both techniques. It is clear that  the
chosen values of the parameters of the correlation integral (middle values)
suffice to obtain good results; the improvements achieved by shifting the values
of the parameters towards better outputs are negligible and at an expensive cost
in numerical work. Instead, shifting the parameters to the other end do have
influence in the output. We conclude that the proposed values of the parameters
used in obtaining the correlation integral are good enough also in cases of
potentials that include chaotic regions.

\begin{table}
 \centering
  \caption{Comparison of different values of the parameters of the correlation
integral method, using orbits integrated in the H\'enon--Heiles potential, 
and the SALI as a gauge. In each row, the value of the corresponding parameter
is shifted, while mantaining the value of the other two at their preferred
(middle) values.}
\label{tabla3}
  \begin{tabular}{@{}lccc@{}}
  \hline
   Parameter     &  \multicolumn{3}{c}{Percentages of coincidence}\\
 \hline
$N_{\rm p}=10^4, 3\times10^4, 5\times10^4 $ & 92.0 & 97.1 & 97.8 \\
$\Delta t=0.1, 0.3, 0.5$                    & 91.8 & 97.1 & 97.9 \\
$\varepsilon =10^{-4}, 10^{-5}, 10^{-6}$    & 97.0 & 97.1 & 97.3 \\
\hline
\end{tabular}
\end{table}

\section{Experiments and results}

\begin{figure*}
\epsfxsize=480pt\epsfbox{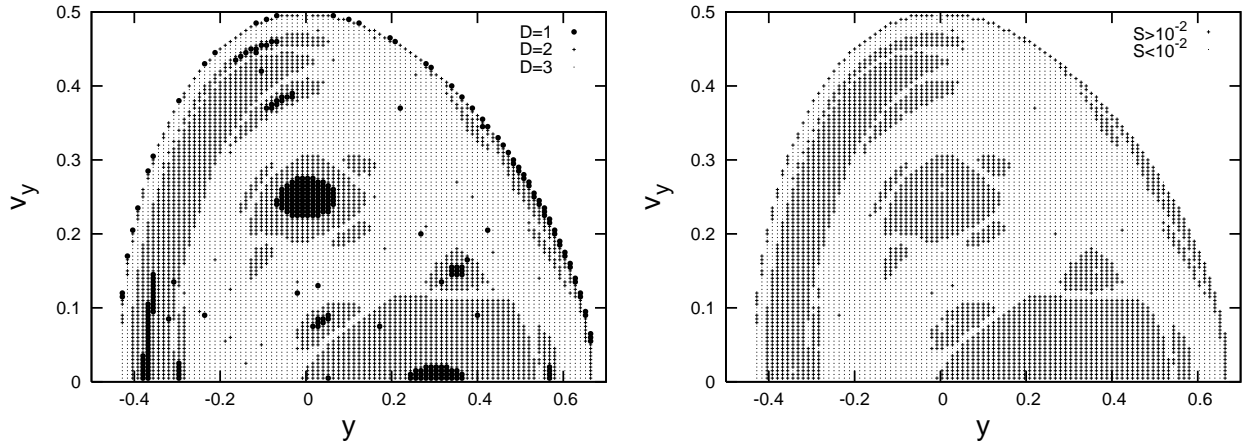}
\caption{Surface of section of the H\'enon--Heiles potential at $E=0.125$,
computed using the correlation integral (left) and the SALI (right).
Each symbol represents the initial point of an orbit that was integrated and
classified separately.} 
\label{henon}
\end{figure*}

\begin{figure}
\epsfxsize=240pt\epsfbox{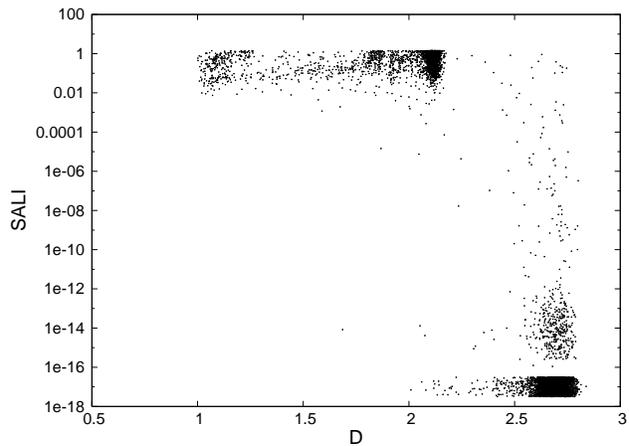}
\caption{Comparison between the values obtained with the correlation integral
and the SALI, for the sample of orbits of Fig. \ref{henon}.} 
\label{civssali}
\end{figure}

\begin{figure}
\epsfxsize=240pt\epsfbox{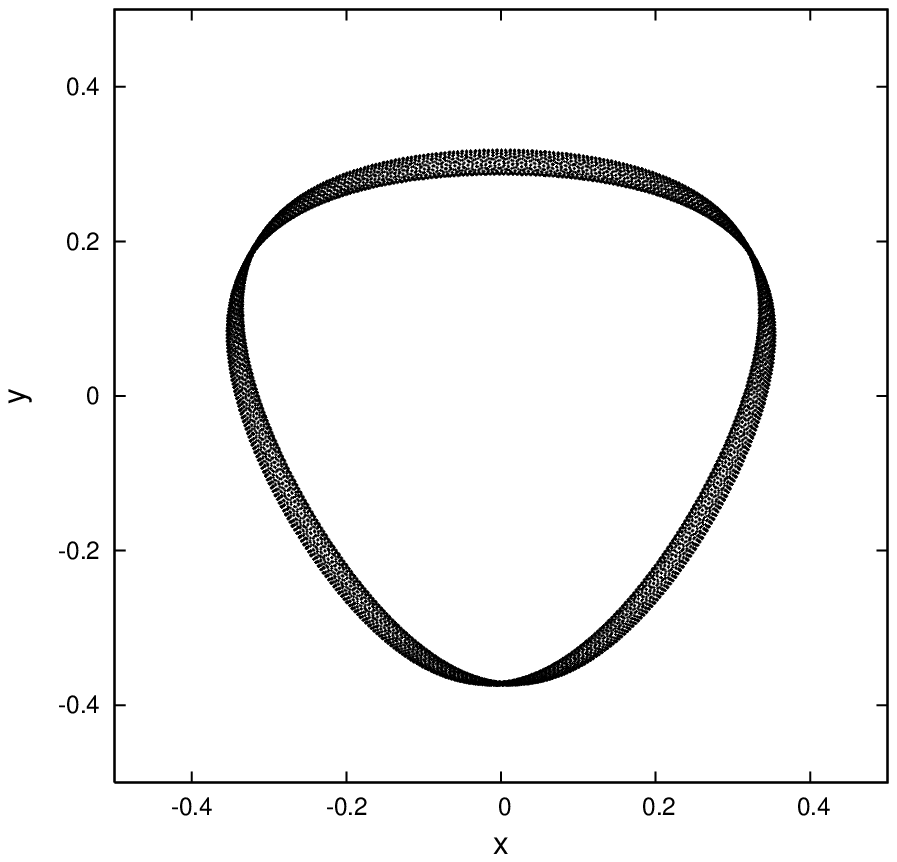}
\caption{Orbit integrated in the H\'enon-Heiles potential with initial
conditions $x=0, y=0.304, v_y=0.05$ and $v_x$ so that $E=0.125$.} 
\label{orbi6801}
\end{figure}

\begin{figure}
\epsfxsize=240pt\epsfbox{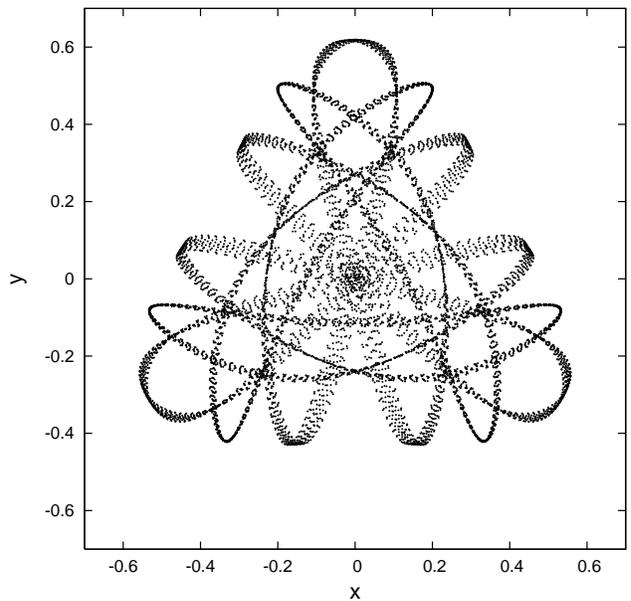}
\caption{Orbit integrated in the H\'enon-Heiles potential with initial
conditions $x=0, y=-0.02, v_y=0.12$ and $v_x$ so that $E=0.125$.} 
\label{orbi4024}
\end{figure}

We have applied the method to several 2D and 3D potentials previously studied in
the literature. Starting from 2D, Fig. \ref{henon} shows the $(x=0,y,v_x>0,v_y)$
surface of section of the H\'enon-Heiles potential (Eq. (\ref{phenon})) at an
energy $E=0.125$, computed both by means of the correlation integral (left) and
by the SALI (right). In the first case, regular orbits with $D=1$ and $D=2$ are
shown with different symbols. In the last case, we have integrated the orbits
until $t=5000$ units and we have taken again the value SALI$=10^{-2}$ as the
bounday between regular and chaotic regimes. Each symbol in the figure indicates
the computed dimension of an orbit, the initial conditions of which are the
position $(y,v_y)$ of the symbol in the plane, $x=0$, and the positive value of
$v_x$ needed to yield $E=0.125$. The classification of the orbits as regular or
chaotic is essentially the same in both methods, except for a few orbits. Fig.
\ref{civssali} shows the actual values of $D$ and SALI obtained for the complete
set of analised orbits; values of SALI$=0$ were arbitrarily distributed around
log$_{10}$SALI$=-17$ in order to visualize them in the logarithmic scale. It can
be seen that, although most orbits are clearly separated by both methods, there
is a set of them that received opposite classifications. We found that the
number of orbits that are in this condition depends on the chosen parameters of
both methods: mainly $N_{\rm p}$ and $\Delta t$ for the correlation integral,
and the boundary value of SALI plus the final time of integration in the case of
the SALI algorithm. Regular orbits, on the other hand, have a suspicious
distribution according to the correlation integral; it is not probable at all
that closed orbits are clustered in that manner. Fig. \ref{orbi6801} shows a
typical orbit inside one of the $D=1$ islands of Fig. \ref{henon} obtained with
the correlation integral: it is clear that this method is not able to
distinguish between this kind of slim $2D$ orbit and a real $1D$ orbit. It is
worth noticing that the few regular orbits detected inside the $3D$ sea by the
SALI are classified as $D=1$ by the correlation integral. Fig. \ref{orbi4024}
shows one of them; as before, it is a very narrow orbit seen as $1D$ by the
correlation integral.

\begin{figure}
\epsfxsize=240pt\epsfbox{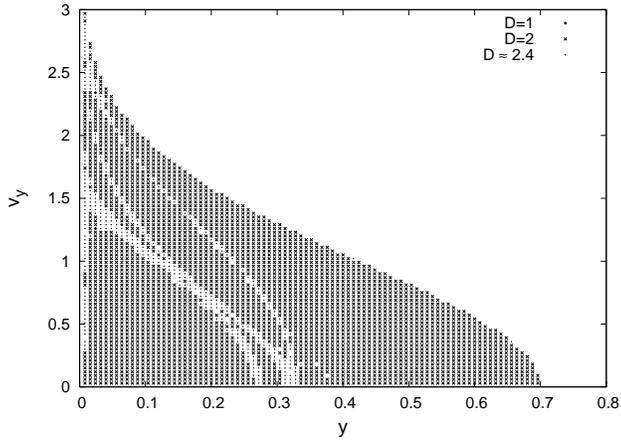}
\caption{Same as Fig. \ref{henon}, but for the logarithmic potential at
$E=0$.}  
\label{log}
\end{figure}

\begin{figure}
\epsfxsize=240pt\epsfbox{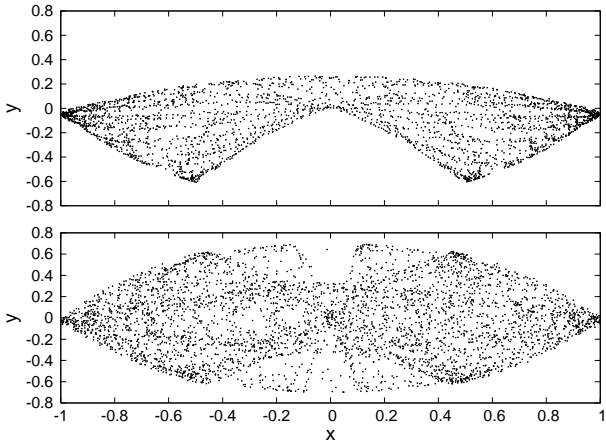}
\caption{Two different time intervals of the same orbit integrated in the
logarithmic potential. The upper panel, between $t=2598$ and $t=6785$, shows
that the orbit is completely regular in this interval. The lower panel, between
$t=40421$ and $t=46195$, shows the orbit transiting between three regular
regimes.}  
\label{trozos}
\end{figure}

\begin{figure}
\epsfxsize=240pt\epsfbox{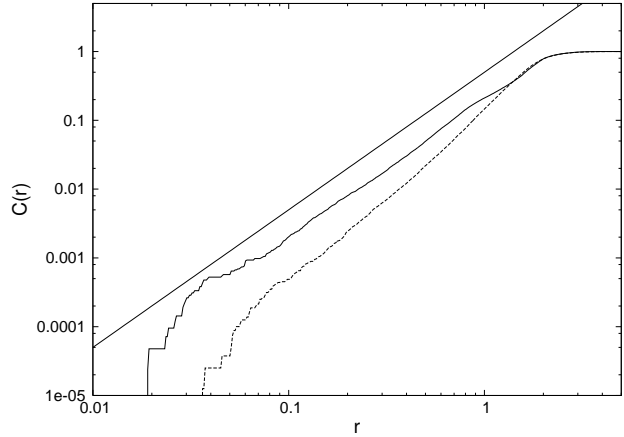}
\caption{Correlation integrals of the portions of orbit shown in Fig.
\ref{trozos}. Solid line: first interval; dashed line: second interval. A
straight line of slope 2 is plotted for reference.}
\label{cder.reca}
\end{figure}

\begin{figure}
\epsfxsize=240pt\epsfbox{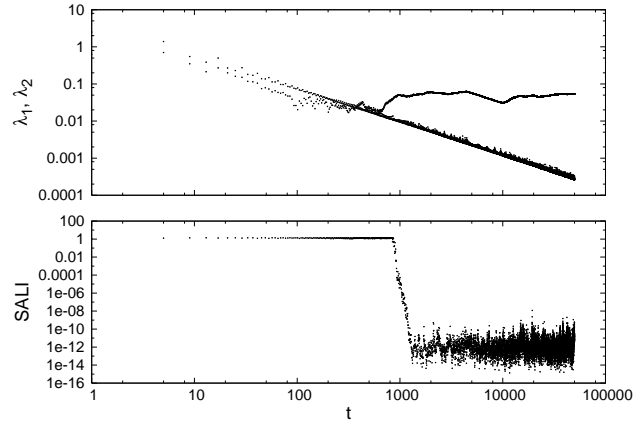}
\caption{Positive Lyapunov exponents and the SALI of the orbit shown in Fig.
\ref{trozos}. The first episode of irregularity occurs near $t=750$, which is
the transition detected by both methods.}
\label{o1334ls}
\end{figure}

\begin{figure}
\epsfxsize=240pt\epsfbox{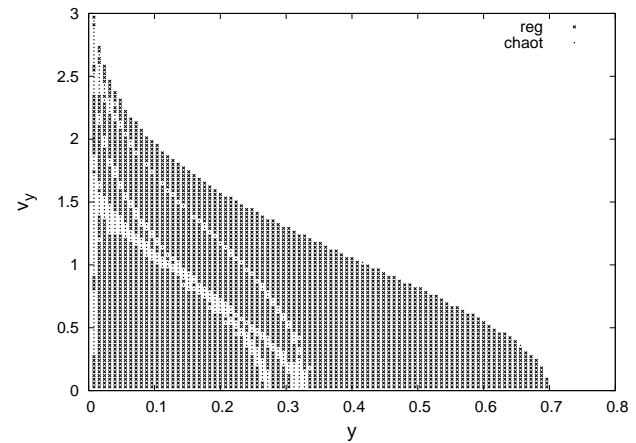}
\caption{Same as Fig. \ref{log}, but computing the chaoticity with Lyapunov
exponents. All the orbits were integrated until $t=10,000$.}
\label{loglya}
\end{figure}

\begin{figure}
\epsfxsize=240pt\epsfbox{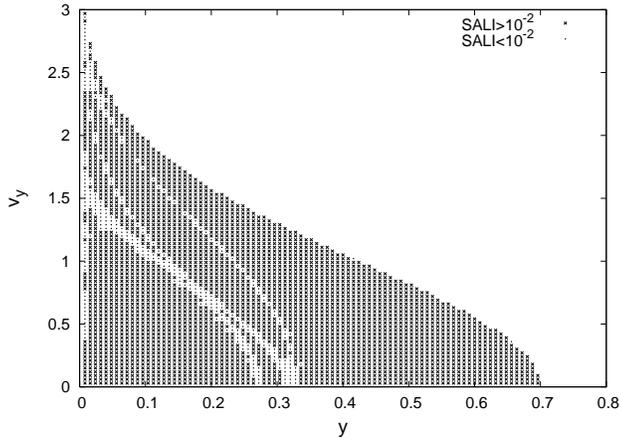}
\caption{Same as Fig. \ref{log}, but computing the chaoticity with the SALI. The
threshold value was chosen at SALI$=10^{-2}$.  All the orbits were integrated
until $t=5,000$.}
\label{logsali}
\end{figure}

We also have classified orbits in the 2D logarithmic potential
\begin{equation}
\Phi_{\rm L}={v_0^2\over 2} \ln \left(R_{\rm c}^2 + x^2 + {y^2\over q^2}
\right),
\end{equation}
where $v_0$, $R_{\rm c}$ and $q$ are constants. Fig. \ref{log} shows the
$(x=0,y,v_x>0,v_y)$ surface of section of the logaritmic potential with $v_0=1$,
$R_{\rm c}=0$, and $q=0.7$, at an energy $E=0$ (a value that in an $R_{\rm
c}=0$, scale free potential, is arbitrary). This can be compared with Fig. 1 of
\citet{ms89} and Fig. 16 of \citet{ca98}, where the orbits were classified using
spectral dynamics. There is an interesting result here: the regions
corresponding to chaotic orbits are filled with orbits that, according to the
results of the correlation integral, move on a manifold of dimension $D=2.4$,
or, equivalently, they obey 1.6 isolating integrals of motion! This of course
cannot be true. We isolated one of these orbits, and studied it in detail, in
order to find out the origin of this behaviour. Fig. \ref{trozos} shows two
portions of this orbit, selected at different time intervals. The upper panel
shows an ordinary regular orbit, whereas the lower panel shows that the orbit is
transiting between different regular regimes. The corresponding correlation
integrals are shown in Fig. \ref{cder.reca}. The orbit is effectively a regular
orbit during the first stage ($D=2$), but its correlation integral takes a slope
$D=2.6$ during the second interval. These regimes were the only two found along
any investigated portion of the orbit. The intervals in which the slope takes
the value $D=2.6$ are those in which the orbit is merely transiting between
different regular regimes. The histograms of the correlation integral of the
regular parts plus those of the transiting parts add up to yield the final
$D=2.4$ figure. That is, the orbit never moves in a 3D manifold, nor it moves
permanently in a 2D manifold. Fig. \ref{o1334ls} shows, for this orbit, the
evolution of its positive Lyapunovs exponents and of the SALI. Here, and in the
rest of this work, Lyapunov exponents were computed using a Gram-Schmidt
orthogonalization of four displacement vectors every time step, renormalizing at
the same time the vectors, following the recipe of \citet{bggs80}. In order to
determine the threshold value between regular and chaotic regimes, we followed
the recipe of \citet{cmvw03}. Both Lyapunov exponents and SALI find this orbit
chaotic, illustrating the fact that it is not moving at all times in a fixed 2D
manifold. This last statement is corroborated in Fig. 16 of \citet{ca98}, in
which all these orbits with slopes $D\simeq 2.4$ in their correlation integrals
were classified as irregular by the spectral dynamics. Moreover, as shown in
Figs. \ref{loglya} and \ref{logsali}, all these orbits are classified as chaotic
according to their Lyapunov exponents or SALI values, i.e., they all show
sensitivity to the initial conditions. Therefore, we have here a numerical proof
that, in a 2D potential, a chaotic orbit (defined as having sensitivity to the
initial conditions) or a 2D irregular orbit (defined as not being regular) is
not neccesarily an orbit that moves in a manifold of dimension 3.

\begin{figure}
\epsfxsize=240pt\epsfbox{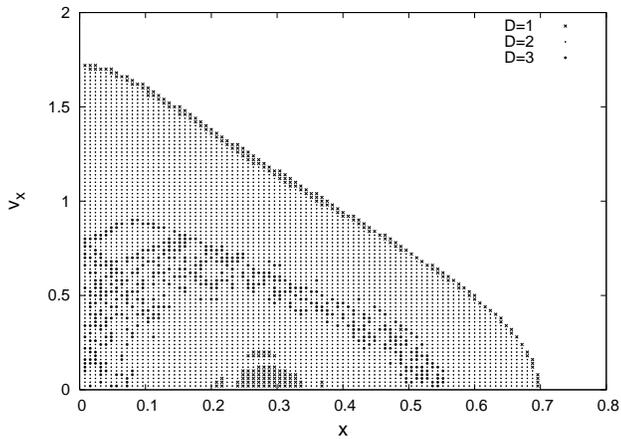}
\caption{Orbital content of the Binney potential with $v_0=1$, $q=0.9$, $R_{\rm
c}=0.14$ and $R_{\rm e}=3$, at an energy $E=-0.4641$.} 
\label{binney}
\end{figure}

We also classified orbits in the Binney potential \citep{b82} 
\begin{equation}
\Phi_{\rm B}={v_0^2\over 2} \ln\left[ R_{\rm c}^2 + x^2 + {y^2\over q^2}
-{(x^2+y^2)^{1/2} (x^2-y^2)\over R_{\rm e}}\right],
\end{equation}
where $v_0$, $q$, $R_{\rm c}$ and $R_{\rm e}$ are constants. Fig. \ref{binney}
shows the $(x,y=0,v_x,v_y>0)$ surface of section of this potential, with
$v_0=1$, $q=0.9$, $R_{\rm c}=0.14$ and $R_{\rm e}=3$, at an energy $E=-0.4641$.
This figure may be compared with Figs. 3.41 and 3.42 of \citet{bt08}, where 
several orbits puncturing this surface of section are showed; the chaotic sea
and the regular regions and islands are well reproduced. The two regions of
$D=1$ orbits are composed, again, of orbits near closeness, which the
correlation integral sees as unidimensional.

We have also classified a set of 3472 orbits previously analised by
\citet{mcw05}, integrated in an analytical potential obtained from a cold
collapse of 100,000 particles (see \citet{mcw05} for details). They classified
the orbits by computing their Lyapunov exponents, and the resulting regular ones
were further classified by means of the spectral dynamics. The dimension
obtained with the correlation integral was rounded to the nearest integer in
order to compare both methods. Table \ref{tabla2} shows the results.

\begin{table}
\centering
\caption{Comparison between the dimension of the manifold obtained using
Lyapunov exponents plus spectral dynamics (rows) and using the correlation
integral (columns), for 3472 orbits integrated in the potential described in
\citet{mcw05}.}
\label{tabla2}
\begin{tabular}{@{}rrrrrr@{}}
\hline
D & 1 & 2   & 3    & 4   & 5    \\
1 & 0 & 0   & 0    & 0   & 0    \\
2 & 0 & 10  & 78   & 2   & 1    \\
3 & 0 & 137 & 1406 & 6   & 4    \\
4 & 0 & 11  & 235  & 38  & 29   \\
5 & 0 & 8   & 236  & 214 & 1057 \\
\hline
\end{tabular}
\end{table}

Inspecting the table, it is quite clear that the results are not the same in
both classifications. In order to quantify this, we performed a crosstabulation
analysis \citep{ptvf94}, using Table \ref{tabla2} as a contingency table. Using
Sakoda's adjusted Contingency Coefficient $V$ as an indicator of the strength of
association between the results of both methods, the value obtained was
$V=0.48$, an intermediate value of correlation. To measure the significance of
this figure, we performed a $\chi^2$ test, resulting in the probability of
obtaining by chance our value of $\chi^2=2392.4$ with 9 degrees of freedom (note
the null row and the null column) being less than $10^{-6}$, i.e., the value of
the correlation $V=0.48$ is indeed statistically significant, and therefore the
classifications indeed differ one another. Other indicators of association gave
similar results.

\begin{figure}
\epsfxsize=240pt\epsfbox{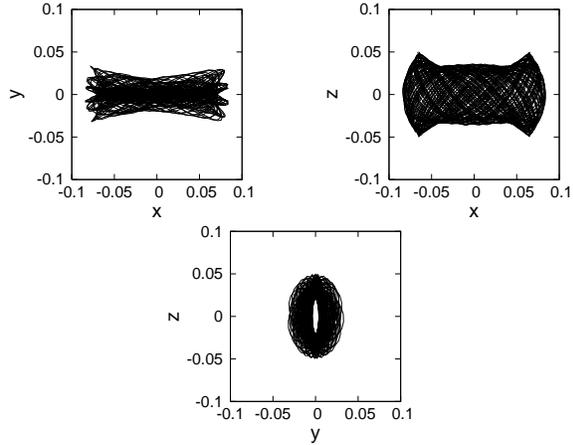}
\caption{Example orbit in configuration space.} 
\label{orb}
\end{figure}

\begin{figure}
\epsfxsize=240pt\epsfbox{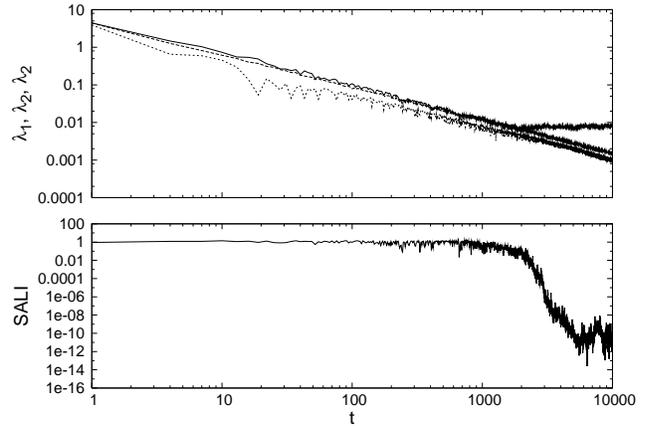}
\caption{The three positive Lyapunov exponents and the SALI, as a function of
time, corresponding to the orbit of Fig. \ref{orb}.} 
\label{salya40}
\end{figure}

We analysed in detail a randomly chosen orbit that had different dimensions
according to both methods. Fig. \ref{orb} shows the chosen orbit in
configuration space. According to the correlation integral, this orbit moves in
a manifold of dimension $D=3$ (a regular orbit), but its Lyapunov exponents
render it as a partially chaotic orbit, i.e., $D=4$, as shown in the upper panel
of Fig. \ref{salya40}. On the other hand, the spectral analysis, besides finding
that it is an $x$--tube orbit, yielded $D=3$. (We recall that only those orbits
classified as regular by their Lyapunov exponents were further classified using
the spectral dynamics; thus, this orbit appears in the table with $D=4$ and not
with the $D=3$ that the last method would have assigned to it.) Also, the lower
panel of Fig. \ref{salya40} shows that, although the value of the SALI does not
tend to the numerical zero ($\simeq 10^{-16}$ in our double precision
experiments), it does have an enough low value to allow classifying the orbit as
chaotic. Visually inspecting Fig. \ref{orb}, it appears to be indeed a regular
orbit; however, a closer inpection reveals a slight dishevelled aspect, which is
in general the signature of a sticky orbit, i.e., an orbit that is irregular but
that wanders during a long time close to a regular torus. This result shows the
limitations of the correlation integral to cope with sticky orbits. On the other
hand, sticky orbits may remain confined to a definite volume of the phase space
during large periods of time, therefore lacking the characteristic diffusion of
other chaotic orbits through phase space that may greatly influence the global
dynamics of an entire stellar system. If the period of stickiness is larger than
the period of interest, the sticky orbit can be considered effectively regular,
as the correlation integral and the spectral dynamics do; in other
circumstances, it should be classified as chaotic, as the Lyapunov exponents or
the SALI do. This analysis enhances the fact that in order to understand the
nature of some difficult orbits we should combine information from several
techniques.

We have examined some other orbits with $D=3$ according to the correlation
integral, and with $D>3$ according to the Lyapunov exponents. Most of them
followed the same pattern as the last example; however, in a few cases, the
orbits were clearly irregular, and therefore correctly classified by the
Lyapunov exponents; the SALI classified these orbits as chaotic in all the
cases. Orbits with $D=4$ according to one method and $D=5$ according to the
other were not analized, because their true dimensions cannot be asserted
visually nor in other independent ways.

\begin{figure}
\epsfxsize=240pt\epsfbox{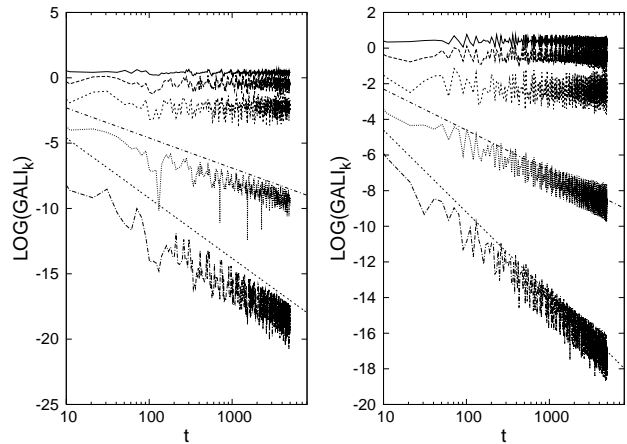}
\caption{GALI$_k(t)$ for the example orbits. From top to bottom: GALI$_2$,
GALI$_3$, GALI$_4$, GALI$_5$ and GALI$_6$. The straight lines have slopes $-1$
and $-2$.} 
\label{galiso}
\end{figure}

With respect to the regular orbits, a possible source of the differences between
classifications might be the rounding of the dimension obtained with the
correlation integral to the nearest integer. We shifted the limit between
integers, and found that, for example, varying the boundary between orbits with
$D=2$ and with $D=3$ in the interval $D\in [2.3,2.7]$ barely improves the
coincidence with the classification of the spectral dynamics. We therefore took
one $D=2$ orbit according to the correlation integral, but classified $D=3$ by
the spectral dynamics, and computed its GALI$_k$ indices in order to determine
its dimension. We expect GALI$_k$ behave as indicated in the second and third
columns of Table \ref{galis} if the orbit moves on a 2D or 3D torus,
respectively \citep{sba08}. The left panel of Fig. \ref{galiso} shows the
computed values of the GALI$_k$ indices for this orbit; the fourth column of
Table \ref{galis} reproduces this result in terms of powers ot time. As can be
seen, the behaviour is not what one would expect. A similar analysis using
another orbit for which the correlation integral yielded $D=3$ but classified as
$D=2$ by the spectral dynamics gave exactly the same behaviour in its GALI$_k$
indices (Fig. \ref{galiso}, right panel). There is the possibility that one or
more of our initial deviation vectors to compute the indices were tangent to the
respective tori. But, even in this case, the expected power laws \citep{sba07}
would not coincide with those computed. Evidently, this result deserves a deeper
study, but which is beyond the scope of this work.

\begin{table}
\centering
\caption{Expected and computed behaviours of the GALI$_k$ indices.}
\label{galis}
\begin{tabular}{@{}rrrrrr@{}}
\hline
                & Expected if $D=2$ & Expected if $D=3$  & Computed \\
\hline
GALI$_2\propto$ & $t^0$             & $t^0$              & $t^0$    \\
GALI$_3\propto$ & $t^{-1}$          & $t^0$              & $t^0$    \\
GALI$_4\propto$ & $t^{-2}$          & $t^{-2}$           & $t^0$    \\
GALI$_5\propto$ & $t^{-4}$          & $t^{-4}$           & $t^{-1}$ \\
GALI$_6\propto$ & $t^{-6}$          & $t^{-6}$           & $t^{-2}$ \\
\hline
\end{tabular}
\end{table}

Therefore, we turn to a visual inspection of some of these orbits. Although we
found, as was the case with the chaotic orbits, that there were a few orbits
misclassified by the correlation integral that were correctly classified by the
spectral dynamics, we found that most of them seemed to have the dimension given
by the correlation integral. When examining their Fourier spectra to find out
why the spectral dynamics method assigned these orbits a wrong dimension, we
found that those spectra had close lines, which is the single most important
numerical problem of the spectral dynamics. 

\section{Conclusions}

We have analized a method to find the dimension of the manifold on which an
orbit moves, dubbed the correlation integral. This amounts to find out how many
isolating integrals of motion the orbit has. In turn, this last number allows to
classify the orbit as regular or chaotic, and, among these two categories,
whether it is closed or resonant, of whether it is partially or fully chaotic.
The method turns out to be easy to implement, but it depends on a number of
numerical parameters which have to be chosen with some care in order to obtain
good results. We have analized the most important parameters, finding the
numerical values that allow the method to be reliable. 

The method was applied to orbits integrated in a St\"ackel potential, obtaining
that most of them move in a manifold of dimension 3, as expected. However, a few
orbits with computed dimensions below 3 can be well classified provided that
they are integrated during longer times, as is the case of the computation of
chaoticity by means of Lyapunov exponents. The method was also applied to a
number of other potentials previously studied in the literature, and compared
against other gauges of chaoticity and/or regularity (Lyapunov exponents,
SALI/GALI and spectral dynamics), giving in general satisfactory results.
Detailed analyses of the orbits that were variously classified by the different
methods showed a limitation of the correlation integral, in particular to cope
with near closed orbits, for which $D=1$ is obtained instead of the correct
$D=2$, and sticky orbits, which are already difficult to cope with for any
algorithm intending to classify them. As said before, these results expose the
need of combininig information from several techniques before a conclusive
answer could be given for any particular orbit. For chaotic orbits, in
particular, it was found that the Lyapunov exponents may not give the true
dimension of the manifold on which the orbit moves. The fact that the
exponential divergence may not be a direct measure of the dimension of the
manifold of the trajectory was already proved for maps \citep[see, e.g.,][\S
4.6]{j91}. The experiments described here show that this may be true also for
continuous differential equations. 

A {\sc FORTRAN 77} program that computes the correlation integral is freely
available upon request.

\section*{Acknowledgments}

We would like to thank the referee for very useful comments which allowed to
improve the paper. This work was supported by grants of the Universidad Nacional
de La Plata, the Consejo Nacional de Investigaciones Cient\'{\i}ficas y
Tecnol\'ogicas, and the Agencia Nacional de Promoci\'on Cient\'{\i}fica y
Tecnol\'ogica de la Rep\'ublica Argentina.

\label{lastpage}

\end{document}